\documentclass[twoside]{article}

\usepackage{graphicx}
\usepackage{amsmath}

\usepackage[preprint]{cmpj2e}
\newcommand{\dd}[0]{\mathrm{d}}
\newcommand{\bb}[0]{\begin{eqnarray}}
\newcommand{\ee}[0]{\end{eqnarray}}
\newcommand{\nn}{\nonumber}

\newcommand{\sn}[1]{S_{#1}}
\newcommand{\PF}[0]{\mathcal{Z}}

\newcommand{\prodd}[2]{\overrightarrow{\prod_{#1}^{#2}}}
\newcommand{\prodg}[2]{\overleftarrow{\prod_{#1}^{#2}}}

\newcommand{\action}{\mathcal{S}}
\newcommand{\Sising}{S_{\mathrm{Ising}}}
\newcommand{\Sint}{S_{\mathrm{int}}}

\newcommand{\vk}{{\bf{k}}}
\newcommand{\vp}{{\bf{r}}}
\newcommand{\vq}{{\bf{q}}}
\newcommand{\bet}{\alpha'}
\newcommand{\Cst}{u_0}
\newcommand{\opx}[1]{\mathcal{O}_{#1}}
\newcommand{\opy}[1]{\mathcal{P}_{#1}}
\newcommand{\baropx}[1]{\bar{\mathcal{O}}_{#1}}
\newcommand{\baropy}[1]{\bar{\mathcal{P}}_{#1}}
\newcommand{\coefk}[1]{\alpha_{#1}}
\newcommand{\sgn}[1]{\sigma({#1})}
\newcommand{\pol}{u}
\newcommand{\mass}{m_{\rm BC}}
\newcommand{\massS}{m_{\rm S}}
\newcommand{\masse}{m}
\newcommand{\ch}{\textrm{ch}}
\newcommand{\sh}{\textrm{sh}}

\newcommand{\TR}{{\rm{Tr}}}

\newcommand{\nam}{\Phi}
\newcommand{\mtr}[4]{\begin{pmatrix} #1 & #2 \\ #3 & #4 \end{pmatrix}}

\title[]{Grassmann techniques applied to classical spin systems}
\author[MC-JYF]{M. Clusel\refaddr{label1}, J.-Y. Fortin\refaddr{label2}, }
\addresses{
\addr{label1} Department of Physics and Center for Soft Matter
Research, New-York University, 4 Washington place, New-York NY 10003, USA
\addr{label2} Institut Jean Lamour
D\'epartement de Physique de la Mati\`ere et des Mat\'eriaux,
Groupe de Physique Statistique CNRS - Nancy-Universit\'e BP 70239
F-54506 Vandoeuvre les Nancy Cedex, France
}
\begin{document}

\maketitle

\begin{abstract}
We review problems involving the use of Grassmann techniques in the field
of classical spin systems in two dimensions. These techniques are useful to
perform exact correspondences between classical spin Hamiltonians and field-theory
fermionic actions. This contributes to a better understanding of critical
behavior of these models in term of non-quadratic effective actions which can been seen as an 
extension of the free fermion Ising model. Within this method, identification of bare masses allows 
for an accurate estimation of critical points or lines and which is supported by Monte-Carlo results
and diagrammatic techniques.
\pacs 02.30.Ik ; 05.50.+q ; 05.70.Fh
\end{abstract}

\section{Introduction}

Classical an quantum spin models such as Ising model play an important role in
the field of statistical physics as they allow for an accurate understanding of 
critical phenomena in general. Many techniques~\cite{Mattis,Schultz}
were developed in order to
deal with the difficulty of estimating the partition function and other 
thermodynamical properties in the critical region in dimension more than one.
An exact mathematical description of the two-dimensional (2D) Ising model relies on 
the Jordan-Wigner transformation~\cite{Jordan} which maps the product of
Boltzmann weights onto a 
fermionic action of free fermions with a mass vanishing at the second order
critical temperature given in dimensionless units $T_c=2/\ln(1+\sqrt{2})\simeq
2.\,2691851$. Also a method based on
the correspondence between the Ising model and dimer problems~\cite{fisher66}
uses
the notion of Pfaffians, which are directly connected to integrals over
Grassmannian objects. Both fermions and Grassmann variables are therefore
closely tied to the Ising model. A direct introduction of Grassmann variables 
as an alternative tool to solve the Ising model was done long ago
in the 80's by Bugrij~\cite{bugrij91} and
Plechko~\cite{Plechko85a} (see also a later discussion by
Nojima~\cite{nojima98}). It is based on a simple integral representation of
the individual Boltzmann weights and which has the property to decouple the
spins. The price to pay is a non-commutativity of terms arising from this
representation. In order to deal with this particular representation, Bugrij
used two families of Grassmann variables which commute with each other, then
identified the resulting functional integral of the partition
function with a determinant. From another point of view, Plechko introduced
symmetries which order the non-commuting quantities so that the sum over the
spins can be performed exactly. 
In this paper we review the process of how to generalize Plechko's method for
Blume-Capel model~\cite{cf08}, which is the simplest model beyond Ising, to
spin-S Ising models and how to construct an exact fermionic action for each
case. This would provide a natural extension of the exact fermionic quadratic
action found for the spin $S=1/2$ Ising model.
In particular, we will build on previous work on the Blume-Capel (BC) case $S=1$
~\cite{cf08} where a
line of second-order critical points is terminated by a tricritical point. This
is the next case beyond the Ising model and which possesses a rich critical
behavior. This model was used to qualitatively explain the phase
transition in a mixture of He$^3$-He$^4$ adsorbed on a 2D surface~\cite{BeG71}.
 Below a 
concentration of 67\% in He$^3$, the mixture undergoes a $\lambda$ transition and the two components
separate through a first order phase transition with only He$^4$ superfluid. 
On a 2D lattice, He atoms are represented by a spin-like variable, according to 
the following rule: an He$^3$ atom is associated to the value 0, whereas a He$^4$ is represented
by a classical Ising spin taking the values $\pm 1$. Within this framework,
all the lattice sites are occupied either by an He$^3$ or He$^4$ atom. 
In addition to nearest-neighbor interactions, the energy includes a term
$\Delta_0 \sum_{mn}S_{mn}^2$, with $S_{mn}^{2}=0,1$, to take into account
a possible change in vacancies number. $\Delta_0$ can be viewed as a
chemical potential for vacancies, or as a parameter of crystal field in a
magnetic interpretation of atomic physics. It would be in particular interesting
to have a fermionic description
of the BC model in order to obtain more information about the kind of interaction fermions
living on the 2D lattice have compare to the Ising free fermion case.

So one of the main question is how to obtain a generic fermionic action for 
a spin-S model and what does this method teaches us for the BC
model in particular. We explain in the next section the main ideas of
this method.

\section{Description of the fermionization for general classical spin-S
models}\label{SpinS}

Let us consider the following Hamiltonian on a 2D lattice of size $L\times L$
\begin{eqnarray}
\label{ham1}
H &=& -\sum_{m=1}^{L}\sum_{n=1}^{L}J\Big[ \sn{mn}\sn{m+1n}
+ \sn{mn}\sn{mn+1}\Big] 
+\Delta_0\sum_{m=1}^{L}
\sum_{n=1}^{L}\sn{mn}^{2},
\end{eqnarray}
where $J$ is the Ising coupling constant and $\Delta_0$ the splitting 
crystal field or represents a chemical potential in the 
BC case. In particular for $\Delta_0$ large and positive, it favors small spin values.
This crystal field can be replaced by any potential $V(\sn{mn}^2)$ depending 
on the square of the local spin.
 Spins $\sn{mn}$ take $2S+1$ values with $\sn{mn}=-S,-S+1,\cdots,S$.
The partition function is the sum over all possible spin configurations
$\PF=\TR_{}\exp(-\beta H)$.
$\PF$ contains products of the Boltzmann
weights $\exp(K\sn{mn}\sn{m+1n})$ (where $\sn{mn}$ and $\sn{m+1n}$ are
neighboring spins and $K=J/k_BT$) which take $q+1=S(S+1)+1$ different values if
$S$ is an integer, and $q+1=(S+1/2)(S+3/2)$ values if $S$ is half-integer. 
Since there are $q+1$ possible values for each Boltzmann weight, we can project
each of them onto a polynomial function of degree $q$ in the variable
$\sn{mn}\sn{m+1n}$:
\bb\label{Boltz}
\exp(K\sn{mn}\sn{m+1n})&=&\sum_{k=0}^{q}\pol_k(\sn{mn}\sn{m+1n})^k
=\Cst\prod_{\alpha=1}^{q}(1+x_{\alpha}\sn{mn}\sn{m+1n}),
\ee
where the $q+1$ constants $\pol_k$ are determined by solving the linear 
system of $q+1$ equations satisfied by the above relation. To see on specific 
examples how it works, let us consider first the Ising case, $S=1/2$. Since $S$ is 
half integer, we have $q=1$. Therefore 
\bb
\exp(K\sn{mn}\sn{m+1n})=\ch(K/4)+4\,\sh(K/4)\sn{mn}\sn{m+1n},\;\;\pol_0=\ch(K/4)
,
\;\;\pol_1=4\,\sh(K/4).
\ee
In the Blume-Capel model, since $S$ is integer, we have $q=2$ and it is 
straightforward to show that 
\bb\nn
\exp(K\sn{mn}\sn{m+1n})&=&1+\sh(K)\sn{mn}\sn{m+1n}+(\ch(K)-1)(\sn{mn}\sn{m+1n}
)^2 ,
\\
\pol_0&=&1,\;\;\pol_1=\sh(K),\;\;\pol_2=\ch(K)-1.
\ee
For $S$ integer the first coefficient $\pol_0$ is always unity, and from
equation (\ref{Boltz}) we can write
\bb
\pol_k=\Cst\sum_{\alpha_1<\alpha_2<\cdots<\alpha_k}x_{\alpha_1}x_{\alpha_2}\cdots x_{\alpha_k},
\;\;1\le k\le q.
\ee
We will set for convenience in the following $\pol_{k\ge q+1}=0$ since the
polynoms are all finite.
Our purpose is to transform the partition function $\PF$ which is a sum over
spin variables into a multiple integral over Grassmann variables. 
For this let us introduce $q$ pairs of Grassmann
variables~\cite{cf08,Plechko85a,Plechko85b}
$(a_{mn}^{\alpha},\bar{a}_{mn}^{\alpha})$
on each site
for the horizontal direction and $q$ additional pairs
$(b_{mn}^{\alpha},\bar{b}_{mn}^{\alpha})$ for the vertical direction. Here
$\alpha$ takes the values 
$1,\dots q$. There are therefore $4q$ Grassmann variables at each site
of the lattice. In particular the Ising model is represented by two pairs of
Grassmann variables per site which can afterward be reduced to only one
pair~\cite{Plechko85a}. For each couple of terms 
\bb
(1+x_{\alpha}\sn{mn}\sn{m+1n})(1+x_{\alpha}\sn{mn}\sn{mn+1})
\ee
appearing in the partition function, we introduce the following integral
representation
%
\bb\nn
1+x_{\alpha}\sn{mn}\sn{m+1n}=
\int d\bar{a}_{mn}^{\alpha}da_{mn}^{\alpha}
{\rm e}^{a_{mn}^{\alpha}\bar{a}_{mn}^{\alpha}}(1+a_{mn}^{\alpha}\sn{mn})
(1+x_{\alpha}\bar{a}_{mn}^{\alpha}\sn{m+1n}),
\\
1+x_{\alpha}\sn{mn}\sn{mn+1}= 
\int d\bar{b}_{mn}^{\alpha}db_{mn}^{\alpha}
{\rm e}^{b_{mn}^{\alpha}\bar{b}_{mn}^{\alpha}}(1+b_{mn}^{\alpha}\sn{mn})
(1+x_{\alpha}\bar{b}_{mn}^{\alpha}\sn{mn+1}).
\ee
%
From the last expression, we introduce the link factors
$A_{mn}^{\alpha}=1+a_{mn}^{\alpha}\sn{mn}$, $\bar{
A}_{m+1n}^{\alpha}=1+x_{\alpha}\bar{a}_{mn}^{\alpha}\sn{m+1n}$,
$B_{mn}^{\alpha}=1+b_{mn}^{\alpha}\sn{mn}$, and $\bar{
B}_{mn+1}^{\alpha}=1+x_{\alpha}\bar{b}_{mn}^{\alpha}\sn{mn+1}$, so that
the partition function can be written as
\bb\nn
\PF=\Cst^{2L^2}
\TR_{\{\sn{},a,b\}}\left[
\prod_{mn}
e^{\Delta\sn{mn}^2}\right.\left.
\times \Big [
\prod_{\alpha}
(A_{mn}^{\alpha}\bar{A}_{m+1n}^{\alpha})
\prod_{\beta}
(B_{mn}^{\beta}\bar{B}_{mn+1}^{\beta})
\Big ]\right],
\ee
where $\Delta=-\beta\Delta_0$. The mixed trace operator introduced in the last
expression is defined by the following sums and integrals:

\bb \nn
\TR_{\{\sn{},a,b\}}[.]= \TR_{\{\sn{}\}}\int\Big
[\prod_{mn,\alpha}d\bar{a}_{mn}^{\alpha}da_{mn}^{\alpha}
d\bar{b}_{mn}^{\alpha}db_{mn}^{\alpha}
\times
e^{a_{mn}^{\alpha}\bar{a}_{mn}^{\alpha}+b_{mn}^{\alpha}\bar{b}_{mn}^{\alpha}}
\Big ][.].
\ee

Inside the integral symbols, the pairs of Grassmannian link factors in brackets
\bb
(A_{mn}^{\alpha}\bar{A}_{m+1n}^{\alpha}), \;\;
(B_{mn}^{\alpha}\bar{B}_{mn+1}^{\alpha})
\ee
can be moved freely with the other terms since they correspond to
commutative scalars after integration. In particular, it is convenient to
rearrange the products over $\alpha$ in order to put together the link factors
of different $\alpha$ with the same site indices $(m,n)$ using the {\it mirror} ordering symmetry
introduced in Plechko's method~\cite{Plechko85a} in the context of the 2D Ising
model, and which is still relevant in the spin-S case:
\bb\nn
\prod_{\alpha=1}^{q} (A_{mn}^{\alpha}\bar{A}_{m+1n}^{\alpha}) &=&
(A_{mn}^{1}\bar A_{m+1n}^{1})\dots
(A_{mn}^{q}\bar A_{m+1n}^{q}),
\\
\nn
&=&(A_{mn}^{1}(A_{mn}^{2}\dots(A_{mn}^{q-1}
(A_{mn}^{q}\bar A_{m+1n}^{q})\bar A_{m+1n}^{q-1})..
\bar A_{m+1n}^{1}),
\\
\nn
&=&
\left(
\prodd{\alpha=1}{q}A_{mn}^{\alpha}\right)\left(\prodg{\alpha=1}{q}\bar{A}_{m+1n}
^{\alpha}
\right),
\ee
where the arrows indicate that the product is ordered, \textit{i.e.} increasing label $\alpha$ in the first product from left to the right and in the second one from right to the left. For convenience, we will use the
notation 
\bb
\opx{mn}=\prodd{\alpha}{}A_{mn}^{\alpha}, \;\;
\baropx{m+1n}=\prodg{\alpha}{}\bar{A}_{m+1n}^{\alpha},
\ee
for objects on the horizontal links and 
\bb\opy{mn}=\prodd{\alpha}{}B_{mn}^{\alpha},\;\;
\baropy{mn+1}=\prodg{\alpha}{}
\bar { B }_{mn+1}^{\alpha}
\ee 
for the ones on vertical links. Then the partition function can be rewritten as
\bb\nn
\PF&=&\Cst^{2L^2}\TR_{\{\sn{} \}}
\int
 \Big [
 \prod_{mn,\alpha}d\bar{a}_{mn}^{\alpha}da_{mn}^{\alpha}
 d\bar{b}_{mn}^{\alpha}db_{mn}^{\alpha}
 e^{a_{mn}^{\alpha}\bar{a}_{mn}^{\alpha}+b_{mn}^{\alpha}\bar{b}_{mn}^{\alpha}}
 \Big ]
 \prod_{mn}
 e^{\Delta\sn{mn}^2}
 (\opx{mn}\baropx{m+1n})(\opy{mn}\baropy{mn+1})
 \\
 &\equiv&\Cst^{2L^2}
\TR_{\{\sn{},a,b\}} \left[
\prod_{mn}e^{\Delta\sn{mn}^2}
(\opx{mn}\baropx{m+1n})(\opy{mn}\baropy{mn+1})
\right].
\ee

At this stage of the algebra, we use the {\it mirror} and {\it associative}
symmetries which were used for solving the Ising model
~\cite{Plechko85a,Plechko85b} and which are still valid here to rearrange the
operators $\opx{}$ and $\opy{}$. In principle boundary terms should be treated
separately in order to obtain the exact finite size partition function depending
on boundary conditions~\cite{Plechko85a} but they are not relevant in the
thermodynamical limit $L\rightarrow\infty$ we are interested in here. Here we
consider instead the simple case of free boundary conditions, and we obtain the
exact expression after rearrangement of the $\opx{}$ and $\opy{}$ operators:
\bb\nn 
\PF=\Cst^{2L^2}
\TR_{\{\sn{},a,b\}} \left[
\prodd{n=1}{L}
\Big (
\prodd{m=1}{L}
e^{\Delta\sn{mn}^2}
\Big (
\baropx{mn}\baropy{mn}\opx{mn}
\Big )
\prodg{m=1}{L}
\opy{mn}
\Big )
\right].
\ee
Now, from this expression, the spins can individually be summed up from
$\sn{Ln}$ to $\sn{1n}$ for any given $n$. We will need to introduce the
following weights $W_{mn}$ which include all the dependence on the
individual spins $\sn{mn}$ 
\bb\nn
W_{mn}=\sum_{\sn{mn}=\pm 1}
e^{\Delta\sn{mn}^2}
\baropx{mn}\baropy{mn}\opx{mn},
\opy{mn},
\\ \label{sum1}
\equiv
\sum_{\sn{mn}=\pm 1}
e^{\Delta\sn{mn}^2}
\prodd{\alpha=1}{4q}\Big (
1+c_{mn}^{\alpha}\sn{mn}
\Big ),
\ee
where we have defined the following $4q$ sets of Grassmann variables
$c^{\alpha}_{mn}$ in the following order:
\bb\nn
c_{mn}^1&=&x_q\bar a_{m-1n}^q,\;c_{mn}^2=x_{q-1}\bar
a_{m-1n}^{q-1},\cdots,\;c_{mn}^{q}=x_1\bar a_{m-1n}^{1},
\\ \nn
c_{mn}^{q+1}&=&x_q\bar b_{mn-1}^q,\;c_{mn}^{q+2}=x_{q-1}\bar
b_{mn-1}^{q-1},\cdots,\;c_{mn}^{2q}=x_1\bar b_{mn-1}^{1},
\\ \nn
c_{mn}^{2q+1}&=&a_{mn}^1,\;c_{mn}^{2q+2}=a_{mn}^2,\cdots,\;c_{mn}^{3q}=a_{mn}^q,
\\ \label{defc}
c_{mn}^{3q+1}&=&b_{mn}^1,\;c_{mn}^{3q+2}=b_{mn}^2,\cdots,\;c_{mn}^{4q}=b_{mn}^q.
\ee
The sum over $\sn{mn}=\pm 1$ in equation (\ref{sum1}) can be performed by
noticing that only products involving an even number of $\sn{mn}$ give a
non-zero contribution. We also define the scalars (we remind that
$\Delta=-\beta\Delta_0$)
\bb
\coefk{k}=\sum_{\sn{mn}=-S}^{S}\sn{mn}^{2k}\exp(\Delta\sn{mn}^2),
\ee
and the ordered products
\bb
q_{mn}^{(k)}=\sum_{\alpha_1<\alpha_2<\cdots<\alpha_k}c_{mn}^{\alpha_1}c_{mn}^
{ \alpha_2}
\cdots c_{mn}^{\alpha_k}, \;\;q_{mn}^{(0)}\equiv 1,
\ee
with $q_{mn}^{(4q)}=c_{mn}^1\cdots c_{mn}^{4q}$ the term of highest degree in
Grassmann variables. Using these quantities, it is easy to show that the partial
Boltzmann weights (\ref{sum1}) are given by
\bb \label{Wmn}
W_{mn}=\sum_{k=0}^{2q}\coefk{k}q_{mn}^{(2k)}.
\ee
Then the fermionic representation of the partition function can be expressed
as a multiple integral over Grassmannian variables only
\bb\label{PFfinal}
\PF=\Cst^{2L^2}
\TR_{\{a,b,\bar{a},\bar{b}\}}\prod_{mn}W_{mn}.
\ee
For small values of $S$, the weights $W_{mn}$ can be \textit{exponentiated}
so that a fermionic action can be defined. Indeed, since the first term
of $W_{mn}$ is the pure scalar $\coefk{0}$ and the others products of pure commutating Grassmannian objects, it is tempting to exponentiate the sum (\ref{Wmn}) to obtain directly a fermionic action. This comes from the simple observation that for any Grassmann variable $a$, we have $1+a=e^a$. Of course, the exponentiation of the sum
(\ref{Wmn}) is more complicate. For example, for commuting objets $a$ and $b$ such as the $q^{k}_{mn}$s, we have $1+a+b=\exp(1+a+b-ab)$. In this case the order of the polynomial object inside the exponential is bigger than in the original sum since
the extra counter-term $ab$ is necessary for the identity to be exact. 
These weights are moreover connected by
nearest-neighbor interactions hidden in the variables $c_{mn}^{\alpha}$. In the
case of the Ising model, where the exponentiation can be done quite easily, the argument of the exponential is purely quadratic in the $c_{mn}^{\alpha}$'s
and therefore the partition function can be integrated out with the use of a
determinant or a Bogoliubov transformation in the Fourier space. 
Moreover, the $4q=4$ Grassmann variables
in this case can be reduced to $2$ by partial integration of non relevant
variables.
In the BC model, the argument is a polynomial of degree 8 in Grassmann
variables since there are 8 independent variables $(4q=8)$. In general we expect
naturally the argument to be at most a polynomial of degree $4q$ in these
variables, which can be 
reduced or not by partial integrations. Except for the case $q=1$ however the
partition function 
can not be expressed as a determinant, so that a full exact solution of the
partition function can not be found this way. If the action is quadratic, the
use of the following Gaussian integral~\cite{berezin}, defined on Grassmann set
of variables
$\{a_i,\,\bar a_i\}_{i=1,..,N}$, and for a square matrix $A$
\bb
\int \prod_{i=1}^{N}\dd \bar a_i \dd a_i \exp\left(\sum_{i,j=1}^{N}
a_iA_{ij} \bar a_j\right)=\det A\,,\;\;
\label{deta1}
\ee
allows us to express the partition function as a determinant.
Quadratic fermionic form in the exponential (\ref{deta1}) is
typically called {\em action} for a free-field theory. When the action is non-quadratic, the integral is not Gaussian and can not be expressed as a determinant,
which yields in principle to a non integrable theory. 
 However, physical information such as $bare$ masses (see last section) can be
extracted from these non-quadratic actions which represent generic theories of
interacting fermions.

\section{Fermionic action of the Blume-Capel model}

In this section, we consider the case $S=1$ (Blume-Capel model) which is the simplest 
example of a classical spin beyond the Ising model. It possesses in the phase
diagram $(T,\Delta_0)$ a second-order critical line separating a ordered phase from a disordered one  and terminated by a tricritical point (see figure \ref{plot} below).
From the previous section equation (\ref{PFfinal}) allows us to write an action after
exponentiation of the Grassmann variables which can be done exactly
after some tedious algebra. The 4 pairs of variables per site can however be
reduced to 2 pairs by partial integration. Another simpler way of obtaining this
BC fermionic action is possible~\cite{cfp08} using the $Z_2$ symmetry of the
spin variables $\sn{mn}$. Indeed the partition function is invariant if we
perform the gauge transformation $\sn{mn}\rightarrow\sigma_{mn}\sn{mn}$ with $\sigma_{mn}=\pm 1$.
In this case it is possible to simplify the process of the previous method and
write an action containing only 2 pairs of variables per site instead of 4:
\bb
\label{PFBC}
\PF&=&(2e^{\Delta}\cosh^2 K)^{L^2}
\int \prod\limits_{m=1}^{L}\prod\limits_{n=1}^{L}
d\bar{a}_{mn}da_{mn}d\bar{b}_{mn}db_{mn}\exp\Big\{\sum\limits_{m=1}^{L}
\sum\limits_{n=1}^{L}
\\ \nn 
& &\Big[\,a_{mn}\bar{a}_{mn} +b_{mn}\bar{b}_{mn}
+a_{mn}b_{mn} +t(\bar{a}_{m-1n} +\bar{b}_{mn-1})
(a_{mn} +b_{mn})
+t^2\,\bar{a}_{m-1n}\bar{b}_{mn-1}
\\ \nn 
&+&\;g_0\;a_{mn}\bar{a}_{mn}b_{mn}\bar{b}_{mn}\,
\exp\,(-\gamma a_{m-1n}\bar{a}_{m-1n}
-\gamma b_{mn-1}\bar{b}_{mn-1}
-t^2\,\bar{a}_{m-1n}\bar{b}_{mn-1})\Big]\Big\}\,,\;\;
\ee
where we have introduced the following constants:
\bb
g_0=\frac{e^{-\Delta}}{2 \cosh^2 K},\;\;\;
\gamma=1-\frac{1}{\cosh K}=1-\sqrt{1-t^2},\;\;\;
t =\tanh K\,.\,
\;\;\;\label{bcint8}
\ee
The fermionic integral (\ref{PFBC}) is the exact expression even for a finite
lattice, provided we assume free boundary conditions for both spins and
fermions. The other possible form for the partition function with
periodic boundary conditions in both direction can be written in a similar
way as the Ising model on a torus~\cite{Plechko85a,liaw99,wuhu02}. The
partition function would be the sum of 4 fermionic integrals with
periodic-antiperiodic boundary conditions for the fermions.
In the expression (\ref{PFBC}), we can recognize the sum of the Ising action,
which here appears as the Gaussian part of the total action
~\cite{Plechko85a,Plechko85b}:
\bb\label{Sising}
\nn \Sising = \sum_{m,n=1}^L a_{mn}\bar{a}_{mn}
+b_{mn}\bar{b}_{mn}+a_{mn}b_{mn}
+t(\bar{a}_{m-1n}
+\bar{b}_{mn-1})(a_{mn}+b_{mn}) +t^2 \bar{a}_{m-1n}
\bar{b}_{mn-1},
\ee 
and a non-quadratic interaction part, which is a polynomial of degree 8 in
Grassmann variables (which can be seen if we expand the exponential inside
the action):
\bb 
\label{Sint} \Sint =g_0 \sum_{m,n=1}^L
a_{mn}\bar{a}_{mn}b_{mn}\bar{b}_{mn} 
\exp
\Big (
-\gamma a_{m-1n}\bar{a}_{m-1n}-\gamma b_{mn-1}
\bar{b}_{mn-1}-t^2\,\bar{a}_{m-1n}\bar{b}_{mn-1}
\Big ).
\ee
This allows us to rewrite the partition function as a fermionic field-theory
in a compact form
\bb
\label{PFFT}
\PF =(2e^{\Delta}\cosh^2 K)^{L^2}\int {\cal D}\bar a
{\cal D}a{\cal D}\bar b {\cal D}b \;\;\exp(\Sising +\Sint)\,.
\ee
The BC model differs from the Ising model by the interaction term in the
action (\ref{Sint}) which is not quadratic. Therefore the BC model is not
solvable in the sense of free fermions as a determinant of some matrix, unlike
the 2D Ising model.

\subsection{Mixed representation of the BC model}

The coupling of Grassmann variables in equation (\ref{Sint}) prevents us to
integrate further and reduce the number of variables per site unlike the Ising
model where the minimal action contains one pair only
~\cite{Plechko99,nojima98}. The minimal action of the Ising model admits
an interpretation in term of Dirac representation of free fermions which 
become massless at the critical point. In a previous work we were able to
reduce the number of Grassmann variables by partially introducing {\it hard
core bosons} in the previous action, since terms such as $\eta_{mn}=a_{mn}\bar
a_{mn}$
or $\tau_{mn}=b_{mn}\bar b_{mn}$ may have an interpretation of local densities
or occupation numbers. Variables $\eta_{mn}$ and $\tau_{mn}$ are commuting and 
nilpotent, $\eta_{mn}^2=\tau_{mn}^2=0$.
We can replace the quantities depending on $a_{mn}\bar a_{mn}$ and $b_{mn}\bar
b_{mn}$, especially in the interaction part, by their respective nilpotent
variables, using, for this task, a general definition of Dirac distribution for
any polynomial function $f$ of $a_{mn}\bar a_{mn}$ or $b_{mn} \bar b_{mn}$
~\cite{cfp08}:
\bb
\nn
f(a_{mn}\bar a_{mn})=\int \dd \eta_{mn}\dd \bar\eta_{mn}
f(\eta_{mn})\exp \left [
\bar\eta_{mn}(\eta_{mn}+a_{mn}\bar a_{mn})\right ],\;\;
\\ 
f(b_{mn}\bar b_{mn})=\int \dd \tau_{mn} \dd \bar\tau_{mn}
f(\tau_{mn})\exp \left [
\bar\tau_{mn}(\tau_{mn}+b_{mn}\bar b_{mn})\right ].\;\;
\label{dirac1}
\ee
A natural definition~\cite{palumbo97} of the integrals involving commuting
nilpotent variables is to impose the following rules (and similar for
$\bar{\eta}_{mn}, \bar{\tau}_{mn}$):
\bb
\int d\eta_{mn}\,(1, \eta_{mn}) =(0,1)\,,\;\;\;
\int d\tau_{mn}\,(1, \tau_{mn}) =(0,1)\,. \;\;\;
\ee
This change of variables allows us now to integrate over the $a_{mn}$'s and
$b_{mn}$'s in the new action. One advantage is that after this operation there
are only two fermionic variables per site, although two additional pairs of
bosonic variables have been introduced. In fact we can integrate over one pair
of bosonic variables~\cite{cfp08}, for example $\bar\eta_{mn},\,\bar\tau_{mn}$,
using the help of integration rules and Dirac function given by (\ref{dirac1}).
At the end, it remains a mixed action made of one pair per site of fermionic and
bosonic variables respectively, with an interaction between
fermions and bosons. A convenient replacement of the variables $\bar
a_{mn}$ by $c_{mn}$ and $\bar b_{mn}$ by $-\bar c_{mn}$ in the final integral
leads us to isolate the minimal local action for the pure Ising model
~\cite{ple95amm,ple98} with one pair of Grassmann variables per site:
\bb
\Sising=c_{mn}\bar c_{mn}
+t(c_{mn}+\bar c_{mn})(c_{m-1n}-\bar c_{mn-1})
-t^2c_{m-1n}\bar c_{mn-1},\;\;
\label{Sising2}
\ee
and the interaction part
\bb
\Sint=g_0\sum_{m,n}\eta_{mn}\tau_{mn}\left
[(1-\gamma\eta_{m-1n})(1-\gamma\tau_{mn-1})
+t^2 c_{m-1n}\bar c_{mn-1}\right ],
\label{SintM}
\ee
with the quantities
\bb\nn
\bar q_{mn}&=&c_{mn}\bar c_{mn} +tc_{mn}(c_{m-1n}-\bar c_{mn-1})
=c_{mn}[\bar c_{mn} +t(c_{m-1n}-\bar c_{mn-1})]\,,
\\
q_{mn}&=&c_{mn}\bar c_{mn} +t\bar c_{mn}(c_{m-1n}-\bar c_{mn-1})
=[c_{mn} -t(c_{m-1n}-\bar c_{mn-1})]\bar{c}_{mn}\,.\;\;
\label{defq}\;\;
\ee
The Ising part is the same action that results from the integration over
$a_{mn},b_{mn}$ from the original Ising case. The introduction of nilpotent
variables was necessary to achieve this partial extraction of the Ising
contribution. The physical interpretation of the previous mixed representation 
is that it can be possible to describe the BC model with fermionic variables for
the states $S=\pm 1$ and bosonic ones for states $S=0$. In the limit
$\Delta_0 \rightarrow -\infty$, the system is completely described in terms of
fermions (Ising sector), while when $\Delta_0$ is increasing fermions and bosons
begin to interact. Beyond a critical value of $\Delta_{0}$, fermions form
bosonic pairs and in the limit $\Delta_0 \rightarrow + \infty$, all fermions
condense into bosons, leading to a purely bosonic system. This view should be 
supported by further analysis.

\subsection{Corrections to the effective action in the continuum limit}

\begin{figure}[tt!]
\begin{center}
\includegraphics[width=0.8\textwidth]{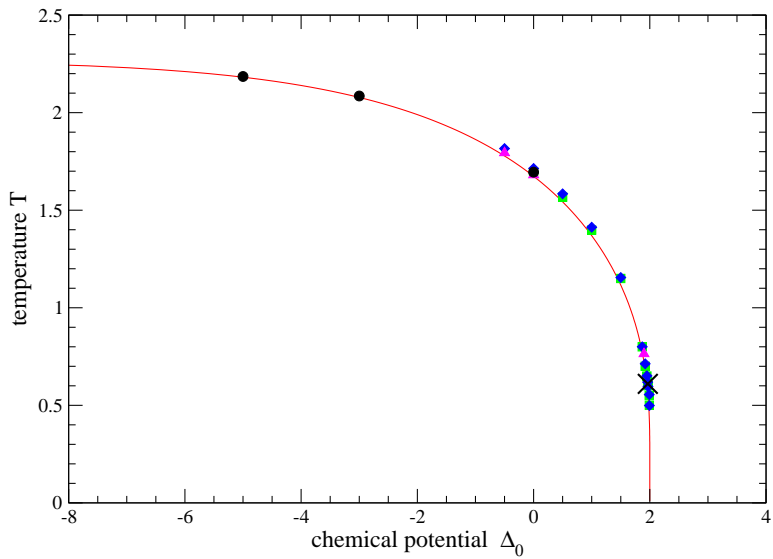}
\unitlength=1cm
\caption{\label{plot} (color online) Figure taken from reference~\cite{cfp08}
showing comparison between critical line defined by the vanishing mass
(\ref{BCmass}) 
(plain red line) and numerical results from Monte
Carlo simulations. The black filled dots are from figure 1, da Silva
\textit{et al.}~\cite{Silva02} (Wang-Landau method). The cross symbol
indicates the tricritical point identified by~\cite{cfp08} using a
Hartree-Fock-Bogoliubov
approximation for the quartic part (\ref{FIint1}) of the effective action. The blue
diamond symbols are from Ref.~\cite{silva06}, the magenta triangles from
Ref.~\cite{xalap98}, and the green squares from Ref.~\cite{beale86}
(see also Table 1 for other numerical values at $\Delta_0=0$).}
\end{center}
\end{figure}

The integration of the previous action (\ref{SintM}) over variables
$(\eta_{mn},\tau_{mn})$ can be performed perturbatively, as part of
an expansion in the low momentum limit. We will define formally the derivatives
of Grassmann variables~\cite{Plechko99}, $\partial_x c_{mn}=c_{mn}-c_{m-1n}$
and
$\partial_y c_{mn}=c_{mn}-c_{mn-1}$ in the limit of large $L$.
In this limit and in the Fourier space, the high order derivatives account in
the action for a small contribution in momenta ${\bf k}=2\pi (m,n)/L$, with
$m,n\ll L$ positive integers. We would like to obtain in this limit the non
trivial part of the non-quadratic interaction in term of variables $c_{mn},\bar
c_{mn}$ only. The procedure is described in reference~\cite{cfp08} and based
partially on substitution rules such as 
\bb
\eta_{mn}\tau_{mn}\rightarrow c_{mn} \bar c_{mn}\,,\;\;\;
\eta_{mn}\rightarrow \bar q_{mn}\,,\;\;\;
\tau_{mn}\rightarrow q_{mn}\,.\;\;\;
\label{rules}
\ee
There are unfortunately more complicate terms in the resulting action than by
using the substitution rules alone, such as 
\bb
g_0^2\gamma^2c_{mn}\bar c_{mn}c_{m+1n}\bar c_{m+1n}
c_{mn+1}\bar c_{mn+1},
\label{example}
\ee
but they can be discarded in the approximation scheme above in the sense they
correspond to corrective terms higher than quartic polynomials or quantities of
the order of ${\cal O}(g_0)$ where $g_0$ is the natural parameter of the
expansion. It is exponentially small in the region where $\Delta_0$ is large 
and negative (Ising behavior). At the lowest order we found that the effective
action
(\ref{Sising2},\ref{SintM}) can be approximated by the following expansion with
respect with $g_0$
\bb
\nn
\action_{\mathrm{effective}}
=\Sising+g_0\sum_{m,n}c_{mn}\bar c_{mn}\left[
(1-\gamma\bar q_{m-1n})(1-\gamma q_{mn-1})
+t^2 c_{m-1n}\bar c_{mn-1}\right ]
\\ \label{Seff}
+g_0^2\gamma^2\sum_{m,n} c_{mn}\bar c_{mn}c_{m+1n}\bar c_{m+1n}
c_{mn+1}\bar c_{mn+1} +\ldots\;. \;\;
\ee
From the previous result, it appears to be suitable to express the quadratic and
quartic parts in the Fourier space (in the large but finite $L$ limit), where we
define the following
transformations
\bb
c(\vp)=\frac{1}{L}\sum_{{\vk}}c_{\vk}\exp(i\vk.\vp)\,,\;\;\;\;
\bar c(\vp)=\frac{1}{L}\sum_{{\vk}}\bar c_{\vk}\exp(-i\vk.\vp)\,.\;\;
\label{Four1}
\ee
The Ising part of the action can be written as
\bb
\label{SIsingK} 
\Sising=\sum_{\vk\in S} [\mass +it(t+1)(k_x-k_y)](c_{\vk} \bar c_{\vk}
-c_{-\vk} \bar c_{-\vk}) +2itk_xc_{\vk}c_{-\vk}
+2itk_y\bar c_{\vk}\bar c_{-\vk},
\ee
with $\mass =1-2t-t^2+g_0$ and the quartic term can be express as
\bb
\Sint=g_0\frac{1}{L^2}\sum_{\vk',\vk'',\vq}
V_{\vk'',\vk''-\vq}c_{\vk'}
c_{\vk''}\bar c_{\vk'+\vq}\bar c_{\vk''-\vq},
\label{FIint1}
\ee
with the potential
\bb
\nn
V_{\vk,\vk'}=-\alpha k_xk'_y+\bet (k_xk'_x+k_yk'_y),\\
\alpha=t(t+2\gamma)\,,\;\;\;\; \bet=\gamma(1-t)\,.
\label{FPint1}
\ee
We notice that the bare mass of the theory is given by 
\bb\label{BCmass}
\mass=1-2t-t^2+g_0=m_{{\rm Ising}}+g_0,
\ee
where $m_{{\rm Ising}}=1-2t-t^2$ is the Ising mass which vanishes at
the critical value $\tanh(K_c)=\sqrt{2}-1$ corresponding to the second order
transition point $T_c=2/\ln(1+\sqrt{2})\simeq 2.\,2691851$ in units of $J/k_B$.
In the BC model, the critical temperature is shifted by the 
parameter $g_0$ which depends on the temperature and $\Delta_0$.
The location of the second order critical line goes from the previous Ising
critical value $T_c=2/\log(1+\sqrt{2})$ when $\Delta_0\rightarrow -\infty$ to
the zero temperature point $(T_c=0,\Delta_0=2)$ continuously where the
transition can be proved to be first order by a simple
energetic argument. In figure (\ref{plot}), we have reported the critical line given
by $\mass=0$ and the different numerical results found in literature
~\cite{cfp08}.
In general the agreement is good, which validates the fermionic theory giving a
bare mass vanishing
at locations close to critical point values found by numerical methods. 
The presence of a tricritical point is induced by the interaction
term (\ref{FPint1}) which renders the second order line instable. To see why, let us consider
the infrared limit on the critical line. The spectrum is given by the lowest terms of an expansion of the effective action with respect with kinetic terms, an the contribution to the partition function,
in the Fourier space, is the product of partial integrals $Z_{\vk}$ such as
$Z=\prod_{\vk}Z_{\vk}$, up to the second order in the momentum $\vk$. For the
Ising model and for small momenta, the factors $Z_{\vk}$ are exactly of the form
$(m_{{\rm Ising}} +Ak^2)$, with $A$ a constant equal to $t(1-t^2)$
~\cite{Plechko99}. The coefficient in front of the term $\vk^2$ in factors
$Z_{\vk}$  can by
described physically as a stiffness coefficient. For the Ising model, the stiffness
is always strictly positive even at the critical point. In the BC case, however, we have 
a line of critical points as $\Delta_0$ varies from negative to positive
values up to $\Delta_0=2$. The effect of the interaction potential (\ref{FPint1}) is to 
modify the expression of the stiffness, which now is no more constant but
depends on the angle of the vector $\vk$ and also the temperature and $\Delta_0$
(see reference~\cite{cfp08} for explicit
details). The result is that in the BC case the effective stiffness
coefficient vanishes at some point on the critical line, at a value close to $\Delta_0=2$, 
which indicates eventually the presence of a tricritical point.
It can be shown that the partition function can be indeed written as a product over the Fourier
modes $Z=\prod_{\vk\in S}Z_{\vk}$ with
\bb
\label{fpFourier}
Z_{\vk}=\mass^2+k^2[A+B\sin 2\theta_k],
\ee
$\theta_k$ being the angle of the vector $\vk$, and $A$ and $B$ depending on temperature
and $\Delta_0$. As long as $|A|$ is larger than $|B|$ on the critical line, the transition is
second order. A singular point can be reached if $A^2=B^2$, in such case $Z_{\vk}$ are not 
all strictly positive if $\mass=0$. Beyond this point the effective action
(\ref{Seff}) is not sufficient to
describe the critical properties of the model. If we compare the fermionic
description of the BC model
to a bosonic Ginsburg-Landau $\Phi^6$ theory describing first order transitions, the presence
of a tricritical point would be equivalent to the fact that both coefficients
of $\Phi^2$ and $\Phi^4$ terms vanish.

\subsection{Critical behavior of the BC model: diagrammatic expansion}

In this section, we further analyze the influence of the interaction
potential $V_{\vk,\vk'}$ on the renormalized mass, in particular the shift of the
critical temperature which was in reference~\cite{cfp08} assumed to be given by
the point where the bare mass $\mass$ vanishes. We would like in particular to apply 
diagrammatic expansion of the effective action (\ref{Seff}). For this, it is useful 
to express the Ising part of the action in term of
Nambu-Gorkov representation of the fermions~\cite{mattuck,Tsvelik}, using
the two-component objects
\bb
\nam_{\vk}=
\begin{pmatrix}c_{\vk} \\ \bar c_{-\vk}
\end{pmatrix},\;
\bar \nam_{\vk}=(\bar c_{\vk}, c_{-\vk}).
\ee
Formally, the Green functions can be defined within this representation by
$2\times 2$ matrices
\bb
\hat G(\vk)=\langle \nam_{\vk}\bar \nam_{\vk}\tau_{3}\rangle,
\ee
where $\tau_3$ is the Pauli matrix
\bb
\tau_{3}=\mtr{1}{0}{0}{-1}.
\ee
The unperturbed part of the Green function $\hat G_0$ is evaluted using the
elements of the non diagonal but quadratic Ising action (\ref{SIsingK}):
\bb
\hat G_0(\vk)=\frac{-1}{|m_k|^2+4t^2k_xk_y}
\begin{pmatrix}\bar m_k & 2itk_y \\ 2itk_x &
m_k
\end{pmatrix},
\ee
 where the momentum-dependent mass is defined by $m_k=\mass +it(t+1)(k_x-k_y)$.
The inverse is given by
\bb
\hat G_0^{-1}(\vk)=
\begin{pmatrix}-m_k & 2itk_y \\ 2itk_x &
-\bar m_k
\end{pmatrix}.
\ee
With this representation and the unperturbed Green function, we can write the
Ising part as
\bb
\Sising=\sum_{k\in S}\bar \nam_{\vk}\tau_{3}\hat G_0^{-1}(\vk)
\nam_{\vk},
\ee
where the set $S$ contains half the momenta of the Brillouin zone. It is
defined by the rule that if $\vk\in S$, then $-\vk$ does not belong to $S$.
The interaction part can be put, after some algebra, into the following form
\bb
\Sint=-g_0\frac{1}{4L^2}\sum_{\vk',\vk'',\vq}
\Big (
\bar\nam_{\vk"-\vq}\tau_{3}
\hat V_{\vk"-\vq,\vk"}
\nam_{\vk"}
\Big )
\Big (
\bar\nam_{\vk'+\vq}\tau_{3}
\nam_{\vk'}
\Big ),
\ee
where the sum is not restricted to the ensemble $S$. We define the potential
matrix $\hat V$ by
\bb
\hat V_{\vk"-\vq,\vk"}=\mtr{V_{\vk",\vk"-\vq}}{0}{0}{V_{\vk"-\vq,\vk"}}.
\ee
The two diagonal elements of this matrix are not equal since
$V_{\vk,\vk'}$ is not symmetric by exchange of the two momenta $\vk$ and $\vk'$
except when $\vk=\vk'$.

\begin{figure}[tb]
\centerline{\includegraphics[width=0.35\textwidth]{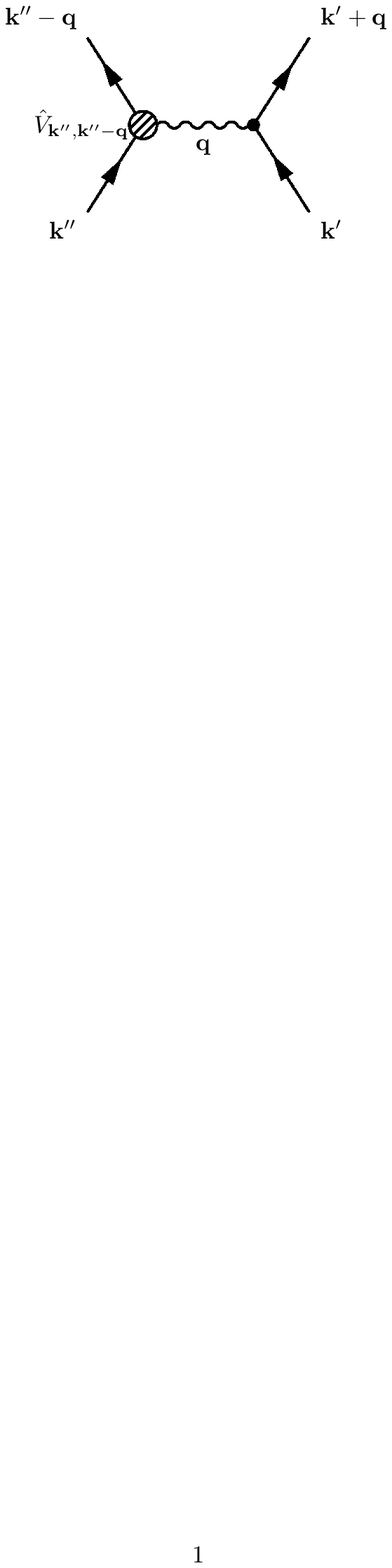}}
\caption{Representation of the interaction part of the potential.
The blob represents the potential interaction with incoming vector
$\vk''$ and outgoing $\vk''-\vq$.}
\label{fig1}
\end{figure}

\begin{figure}[tb]
\centerline{\includegraphics[width=0.65\textwidth]{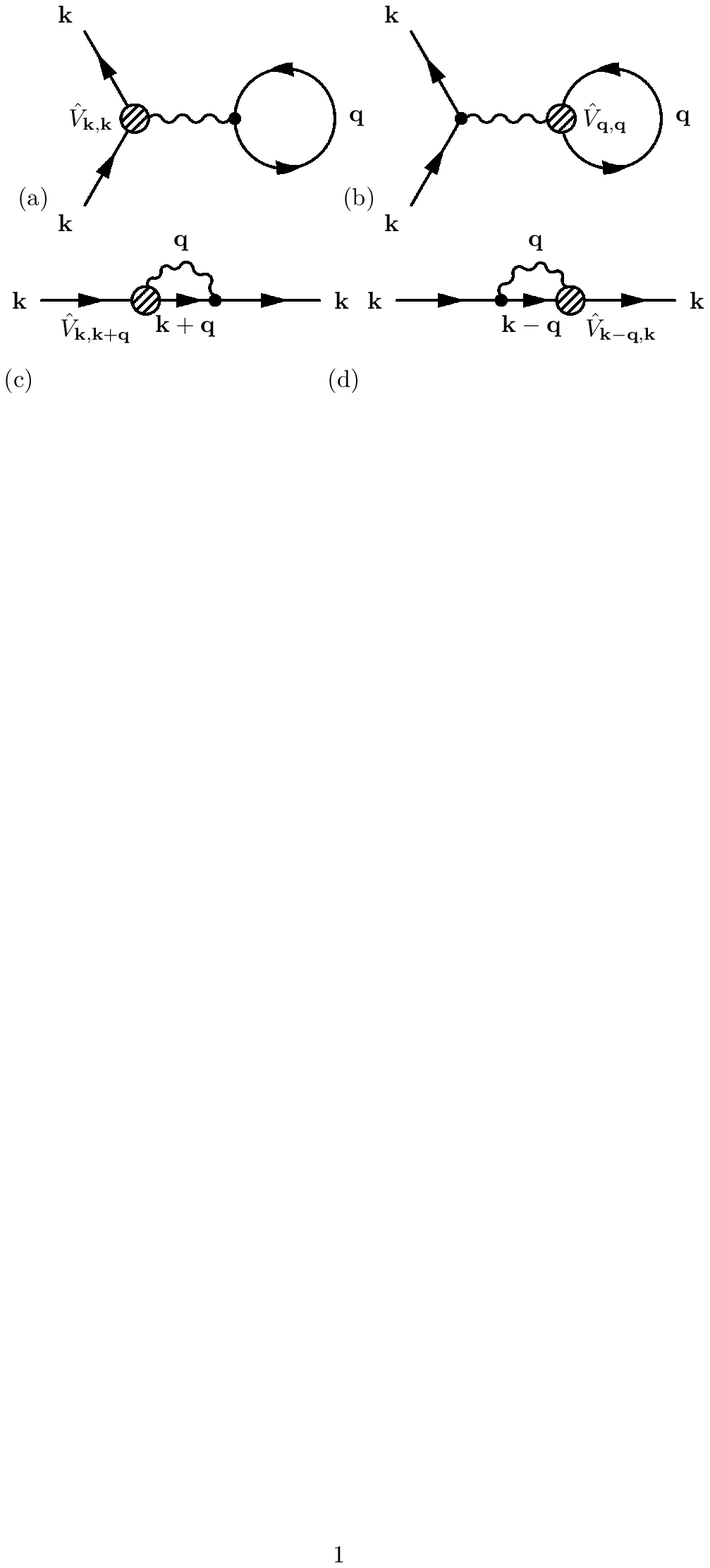}}
\caption{The four diagrams appearing at the lowest order in $g_0$.
Only diagram (b) contributes to the mass in the low momentum limit $k\rightarrow 0$.}
\label{fig2}
\end{figure}

We now perform a diagrammatic expansion with respect with $g_0$ of the
perturbed Green function $\hat G(\vk)$ which will allow us to compute the
corrections to the mass, i.e. the shift of the critical temperature, by
identification of the diagonal elements of the inverse propagator $\hat
\Gamma(\vk)=\hat G(\vk)^{-1}$. The
graphical representation of the matrix potential in term of
diagram is displayed in figure \ref{fig1}. To do so we follow formally the
Feynman rules which lead to the Dyson equation of the inverse-propagator in term
of the self-energy $\hat \Sigma(\vk)$:

\bb
\hat \Gamma(\vk)=\hat G(\vk)^{-1}=\hat G_0(\vk)^{-1}-\hat \Sigma(\vk).
\ee

The first terms contributing to the self-energy are given in figure \ref{fig2}:

\bb\nn
\hat \Sigma(\vk)&=&-g_0\frac{1}{4L^2}\sum_{\vq}\Big [\hat
V_{\vk,\vk+\vq}\hat G_0(\vk+\vq)+\hat G_0(\vk-\vq)\hat V_{\vk-\vq,\vk}
\Big ]
\\
& &-g_0\frac{1}{4L^2}\sum_{\vq}\Big [\hat V_{\vk,\vk}\TR \hat G_0(\vq)+
\hat V_{\vq,\vq}\TR \hat
G_0(\vq)
\Big ].
\ee

The renormalized mass $m_R$ is given in the limit when $k$ is zero by the
diagonal components of the inverse-propagator $\Gamma_{11}(0)=\Gamma_{22}(0)$.
In this limit, only one diagram is not vanishing, which corresponds to the
diagram (b) of figure (\ref{fig2}):
\bb\nn
m_R&=&\Gamma_{11}(0)=\mass +g_0\frac{1}{4L^2}\sum_{\vq}V_{\vq,\vq}\TR \hat
G_0(\vq),
\\ \label{mR} &=&\mass -\mass \times
g_0\frac{1}{2L^2}\sum_{\vq}\frac{V_{\vq,\vq}}{
\mass ^2+t^2(1+t)^2(q_x-q_y)^2+4t^2q_xq_y }.
\ee
The last sum over $q$ can be evaluated in the continuous limit $L\rightarrow
\infty$. Setting $\vq=2\pi(\frac{m}{L},\frac{n}{L})$, we define for large $L$
the two following integrals
\bb\nn
I_1(\mass )&=&\frac{g_0}{8\pi^2}\int_0^{2\pi}\int_0^{2\pi}
dq_xdq_y\frac{q_xq_y}{\mass ^2+t^2(1+t)^2(q_x-q_y)^2+
4t^2q_xq_y},\\
I_2(\mass )&=&\frac{g_0}{8\pi^2}\int_0^{2\pi}\int_0^{2\pi}
dq_xdq_y\frac{q_x^2+q_y^2}{\mass ^2+t^2(1+t)^2(q_x-q_y)^2+
4t^2q_xq_y}.
\ee
These two quantities are finite when $\mass $ vanishes. To see why, we can 
consider polar coordinates $q_x=q\cos\theta$ and $q_y=q\sin\theta$, so that,
near the origin $q=0$ the second integral for example behaves like
\bb
I_2(\mass)\propto\frac{g_0}{8\pi^2t^2(1+t)^2}\int_0\int_0^{2\pi}
q\,dq\,d\theta
\frac{1}{\{\mass/qt(1+t)\}^2+1+\{
-1+2/(1+t)^2 \}\sin(2\theta)}.
\ee
When $\mass=0$, this integral is finite since there is no singularity in the
denominator. Indeed when $\theta=\pm \pi/4$ the modulus of the term
$-1+2/(1+t)^2$ is strictly less than one on the critical line.
This would not be the case if the last term $4t^2q_xq_y$ in the denominator and
coming from the off-diagonal part of the Green function was absent. In this case
the integrals would be singular in the limit of small mass $\mass$,
the denominator would instead be equal to $\{\mass/qt(1+t)\}^2+1-\sin(2\theta)$, 
and the singular part would behave like $1/|\mass |$ by a simple scaling argument, which
would cancel the other mass term $\mass$ in
front of the integrals (\ref{mR}). Then the renormalized mass would be shifted
by a finite quantity, as well as the critical line. Here the renormalization
only concerns the total coefficient of $\mass$ and this does not affect the
critical line location:

\bb
m_R\simeq \mass \Big (
1-\alpha I_1(0)+\alpha' I_2(0)
\Big ).
\ee

\begin{figure}[htb]
\centerline{\includegraphics[width=0.65\textwidth]{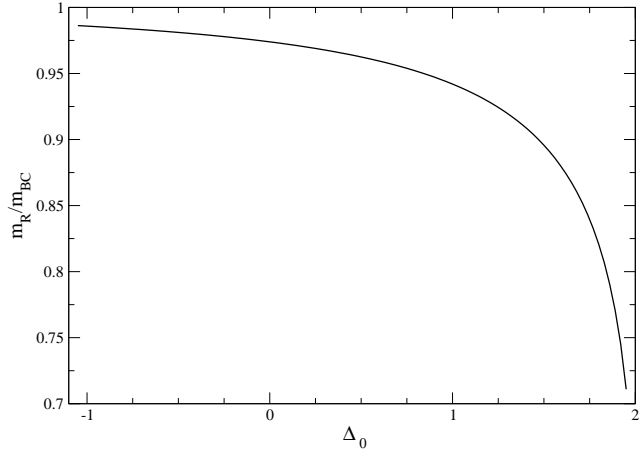}}
\caption{Plot of the coefficient renormalizing the BC bare mass as function
of $\Delta_0$ on the critical line. The values are close to unity except near
the region of the terminating point $\Delta_0=2$ where $g_0$ is not small.}
\label{figmR}
\end{figure}

A plot of the positive ratio $m_R/\mass$ evaluated at criticality $\mass=0$ as
function of $\Delta_0$ is given in figure \ref{figmR}. It is close to unity for
almost all values of $\Delta_0$. 

This analysis shows that corrections to critical temperatures are indeed small
and are of order higher than $g_0$. A further analysis should be carry out to
obtain a finite shift by considering more complex diagrams. It is also supported
by analytical values of $\mass=0$ which are close to numerical results reported
in figure \ref{plot}.

\section{Extension to other spin-S models: generalization of the bare mass}
The previous bare mass computed in the BC model, $\mass $, allowed us to obtain
a precise description of the second-order critical line in the phase diagram.
This was obtained by taking the limit of low momentum in the effective action,
see (\ref{Seff}). This can be generalized for any value of the spin $S$, in
particular for higher values of $S$. The equations obtained in section
\ref{SpinS}, equations (\ref{Wmn}) and (\ref{PFfinal}), are general for
they represent the fermionization of any spins-S model. 

The construction of the fermionic action is however not an easy task, unlike the BC model
which is a simpler case, but we expect to be able to extract a bare mass associated to
non kinetic terms, or term involving derivatives with respect with space variables. At first
approximation, we assume that the partition function and the free energy are singular in the low momentum
limit when this bare mass vanishes.
 In the continuum limit, the $c$'s coefficients defined by relations
(\ref{defc}) can be rewritten using formal derivatives, such as
$c_{mn}^1= x_q(\bar{a}_{mn}^q-\partial_x\bar{a}_{mn}^q)$, etc... The 
derivatives contribute only to the kinetic energy and not to the bare mass.
Keeping the first terms of the expansion, $c_{mn}^1\simeq x_q\bar{a}_{mn}^q$ (as
well as for the other coefficients $c$'s), the weights $W_{mn}$ become
uncoupled in the sense they contain variables depending only on local site $(m,n)$ and 
we define the mass $m_S$ as:
\bb\nn
\massS\equiv\Cst^2\int\Big [
\prod_{\alpha=1}^q
d\bar{a}_{mn}^{\alpha}da_{mn}^{\alpha}
d\bar{b}_{mn}^{\alpha}db_{mn}^{\alpha}
e^{a_{mn}^{\alpha}\bar{a}_{mn}^{\alpha}+b_{mn}^{\alpha}\bar{b}_{mn}^{\alpha}}
\Big ] W_{mn},
\ee
with $c_{mn}^1\simeq x_q\bar{a}_{mn}^q,\cdots,c_{mn}^{q}\simeq
x_1\bar{a}_{mn}^{1}$ and
$c_{mn}^{q+1}\simeq x_q\bar{b}_{mn}^q,\cdots,c_{mn}^{2q}\simeq
x_1\bar{b}_{mn}^{1}$.
 The integral can be evaluated exactly by noticing for example that the
arguments of the exponential 
$b_{mn}^{\alpha}\bar{b}_{mn}^{\alpha}$ can be combined with a
$a_{mn}^{\alpha}(x_{\alpha}\bar{a}_{mn}^{\alpha})$ that appears
in some of the $q^{(2k)}_{k=1\dots 2q}$ products to give a contribution
$x_{\alpha}$. Indeed
using 
the Grassmann integration rules $\int da.a=1$ and $\int da.1=0$, we can write
\bb\nn
& &
\int d\bar{a}_{mn}^{\alpha}da_{mn}^{\alpha}
d\bar{b}_{mn}^{\alpha}db_{mn}^{\alpha}
e^{a_{mn}^{\alpha}\bar{a}_{mn}^{\alpha}+b_{mn}^{\alpha}\bar{b}_{mn}^{\alpha}}
a_{mn}^{\alpha}(x_{\alpha}\bar{a}_{mn}^{\alpha}) =
\\ \nn
& &\int d\bar a_{mn}^{\alpha}da_{mn}^{\alpha}
d\bar b_{mn}^{\alpha}db_{mn}^{\alpha}
\Big( 1+a_{mn}^{\alpha}\bar a_{mn}^{\alpha}+b_{mn}^{\alpha}\bar b_{mn}^{\alpha}
+a_{mn}^{\alpha}\bar a_{mn}^{\alpha}b_{mn}^{\alpha}\bar b_{mn}^{\alpha}
\Big )a_{mn}^{\alpha}(x_{\alpha}\bar a_{mn}^{\alpha})=
\\
\nn
& &
\int d\bar a_{mn}^{\alpha}da_{mn}^{\alpha}
d\bar b_{mn}^{\alpha}db_{mn}^{\alpha}
b_{mn}^{\alpha}\bar b_{mn}^{\alpha}a_{mn}^{\alpha}(x_{\alpha}\bar
a_{mn}^{\alpha})=
x_{\alpha}.
\ee
Also a term $a_{mn}^{\alpha}\bar{a}_{mn}^{\alpha}$ can be combined with
$b_{mn}^{\alpha}(x_{\alpha}\bar{b}_{mn}^{\alpha})$ to give the same
contribution.
Since the $q^{(2k)}$ are ordered, there are also signs to take into
account and coming from the fact the variables $c^{\alpha}_{mn}$ have to
be moved in the correct order before integration. We obtain after some algebra
the general relation
\begin{eqnarray}
\label{resmass}
\massS=\sum_{k=0}^{2q}\coefk{k}R_{k},
\end{eqnarray}
where we have define the following quantities with initial condition $R_0=u_0^2$, 
\begin{eqnarray} 
\label{sumR}
R_k=\sum_{l=0}^{k}\pol_l\pol_{k-l}\sgn{l,k-l},
\end{eqnarray}
and $\sgn{k,l}=1$ if $k$ and $l$ are both even, and $\sgn{k,l}=-1$ otherwise.
We can apply this result to different cases to check the validity of this
relation. For the Ising model ($S=1/2,q=1$) $\pol_0=\ch(K/4)$ and
$\pol_1=4\sh(K/4)$, we obtain $\masse_{1/2}=2\cosh(\Delta/4)(\pol_0^2-\pol_0\pol_1/2-\pol_1^2/16)$,
or
\bb
\masse_{1/2}=2e^{\Delta/4}[1-\sh(K/2)],
\ee
which vanishes at the Ising critical temperature $T_c\simeq 0.567\,296$ 
or with the normalization $t_c\equiv T_c/S^2=2.269\,185$, which is independent,
as 
expected, of $\Delta_0$. 
For the Blume-Capel model ($S=1,q=2$) we have
$\masse_1=1+2\exp(\Delta)(1-2\pol_1+2\pol_2-\pol_1^2-2\pol_1\pol_2+\pol_2^2)$,
or more explicitly
\bb
\masse_1=1+2e^{\Delta}[1-\sh(2K)].
\ee
This mass is directly proportionnal to the mass $\mass$ found in the previous section.
Indeed, we have the relation
\bb 
\mass=g_0\masse_1
\ee
and therefore both masses vanish on the same line of critical points. The coefficient
$g_0$ comes from a global rescaling of the Grassmann variables in the original weights $W_{mn}$ which leads to the
coefficient $g_0^{-L^2}$ in the BC function partition (\ref{PFBC}) and
(\ref{PFFT}), instead of the coefficient $(\Cst)^{2L^2}=1$ in front of
(\ref{PFfinal}). For $\Delta_0=0$ we find in particular that
$t_c=2/\rm{arcsinh}(3/2)\simeq 1.673\,971$ (see table \ref{table1}).

\begin{table}
 \caption{\label{table1}Comparison of critical temperature values at
$\Delta_0=0$, solutions of equation (\ref{genmass}), with other methods
found in bibliography.}
\begin{center}
 \begin{tabular}{c|ccccccc|c}
 Spin S & $S=1/2$ & $S=1$ & $S=3/2$ & $S=2$\\
\hline
$q$   & 1 & 2 & 5 & 6
\\
$T_c$ & $0.567\,296$ & $1.673\,971$ & $3.277\,561$ & $5.351\,248$
\\
$t_c=T_c/S^2$ & $2.269\,185$ & $1.673\,971$ & $1.456\,694$ & $1.337\,812$
\\
Refs. & $2.269$~\cite{Butera03} & $1.689$~\cite{Fox73}, $1.695$~\cite{Silva02} &
$1.461$~\cite{xavier98,Butera03,grandi04} & $1.336$~\cite{Butera03} 
\\
 & & $1.694$~\cite{Butera03}, $1.681$~\cite{xavier98} & 
\\
\hline
 Spin S & $S=5/2$ &  $S=3$ & $S=\infty$\\
\hline
$q$ &  11 & 12 & $\infty$
\\
$T_c$ & $7.890\,888$ & $10.894\,806$ & $\infty$
\\
$t_c=T_c/S^2$ & $1.262\,542$ & $1.210\,534$ & $0.925\,148$
\\
Refs. &  $1.257$ ~\cite{Butera03} & $1.203$ ~\cite{Butera03} & $0.915$
~\cite{Butera03,Bial00}
 \end{tabular}
\end{center}
 \end{table}

The other masses are deduced by iteration of formula (\ref{resmass}) and (\ref{sumR}). 
For higher values of spin $S$, we find:
\bb\nn
\masse_{3/2}&=&2e^{\Delta/4}[1-\sh(K/2)]+2e^{9\Delta/4}[1-\sh(9K/2)]+
2e^{\Delta/4}[1-\sh(K/2)],
\\ \nn
\masse_{2}&=&1+2e^{\Delta}[1-\sh(2K)]+2e^{4\Delta}[1-\sh(8K)],
\\ \nn
\masse_{5/2}&=&2e^{25\Delta/4}[1-\sh(25K/2)]
+2e^{9\Delta/4}[1-\sh(9K/2)]
+2e^{\Delta/4}[1-\sh(K/2)],
\\ \label{massS}
\masse_{3}&=&1+2e^{9\Delta}[1-\sh(18K)]
+2e^{4\Delta}[1-\sh(8K)]
+2e^{\Delta}[1-\sh(2K)].
\ee
For general spin $S$, we can extend the previous results to the formula
\bb
\masse_{S}=\sum_{\sigma=-S}^{S}e^{\Delta\sigma^2}[1-\sh(2\sigma^2 K)].
\label{genmass}
\ee
This is a simple result giving a precise location of the second-order critical
lines by solving the equation of the bare mass $\masse_S=0$. 
We give for comparison tabulated values of $t_c$ at $\Delta_0=0$ in table
\ref{table1} for different values of $S$, and references to numerical results
(Monte-Carlo simulations, high-temperature expansions) given in the literature.
In general, the agreement is good up to $1\%$ in most cases.
For half-integer values of $S$, the models possess an asymptote in the
$(T/S^2,\Delta_0)$ plane. Indeed, for $S=3/2$ for example, the equation given
in (\ref{massS}) predicts the solution
\bb
\Delta_0=-\frac{9t}{8}\log\Big [
-\frac{1-\sh(2/9t)}{1-\sh(2/t)}
\Big ],
\ee
which is bounded by $t_c=2/9\log(1+\sqrt{2})=0.252\,131$ below which there is no continuation
of the second-order critical line.
In the limit of large integer $S$, the model defined in equation (\ref{ham1}) is described by a continuous variable $-1<x_{mn}=\sn{mn}/S<1$ and is called \textit{continuous Ising model}. We can still obtain a finite value of the critical line by taking the asymptotic value of equation (\ref{genmass})
\bb\label{LargeS}
\masse_{S\gg 1}\simeq S\sqrt{2t}\int_0^{\sqrt{2/t}}dx\; e^{-\Delta_0x^2/2}\Big [
1-\sh(x^2) \Big ],
\ee
and in particular for $\Delta_0=0$, we have the following expansion for large $S$
\bb
\masse_S(t,\Delta_0=0)\simeq a(t)S-1+\frac{4}{3tS}+\frac{8}{21t^3S^6}+\cdots
\ee
with $a(t)=2-2\sqrt{2/t}\int_0^{\sqrt{2/t}}\sh(x^2)\,dx$. 
We observe that the rescaled mass $\masse_S/S$ vanishes in this case when
$t_c=0.925\,148$, in good agreement with numerical works for this model
~\cite{Butera03,Bial00}, and it is worth noting that
equation (\ref{LargeS}) also possesses a non trivial solution at $t=0$ which is
simply given by $\Delta_0=4/\sqrt{3}=2.309\,401$. This value is different from
the value $2$ expected for all finite $S$ models~\cite{cf08}. 
It can be suggested that there also exists a tricritical point
before this non physical value is reached.

\section{Conclusion}

In this review paper, we have presented a method which tries to operate a
correspondence
between classical spin models and fermionic systems. We have extended Plecho's
method~\cite{Plechko85a,Plechko85b} based on Ising model to generalized spin-S
systems.
The method is based on the \textit{projection} onto $q$ polynomial components, $q$ depending
specifically on the value of spin $S$, of Boltzmann local weights given by equation (\ref{Boltz}). Then the introduction of $2q$ pairs of 
Grassmann variables per site and the use of special symmetries such as \textit{mirror} and \textit{associative}
symmetries in 2D for Grassmannian objects allows us to perform exactly the sum
over the spin variables.
This gives a representation of spin-S models in term of fermionic multiple integrals (\ref{PFfinal}). Effective
actions can in principle be deduced from this representation. We have shown that such action
can be built exactly for the Blume Capel model $S=1$ (\ref{Seff}) and the bare mass (\ref{BCmass}) gives accurate description 
of the second-order critical line. We have seen that there is no shift of this mass due to the effect of 
quartic potential of the effective theory at the lowest order expansion in the coupling parameter
$g_0$, implying that corrections to the critical temperature may be indeed small. This 
quartic potential is however responsible for the presence of a tricritical point, rendering the second order
line instable by changing the sign of the stiffness coefficient or making the 
free-fermion spectrum itself instable.
For general spin-S model, the bare mass can also be generalized and calculated directly in the
low momentum limit (\ref{resmass}) without knowing the full effective action,
and still gives accurate description of second-order critical points
even in the limit of the continuous Ising model.


\end{document}